\documentclass[aps]{revtex4}
\input{tcilatex}

\begin{document}

\title{Carrier relaxation dynamics in heavy fermion compounds}
\author{J. Demsar$^{\ast }$, L.A. Tracy, R.D. Averitt, S.A.Trugman, J.L.Sarrao, A.J.
Taylor \\
%EndAName
Los Alamos National Laboratory, Mail Stop K764, Los Alamos, NM 87545, USA}

\begin{abstract}
The first femtosecond carrier relaxation dynamics studies in heavy fermion
compounds are presented. The carrier relaxation time shows a dramatic
hundred-fold increase below the Kondo temperature revealing a dramatic
sensitivity to the electronic density of states near the Fermi level.
\end{abstract}

\maketitle

\section{Introduction}

Femtosecond time-resolved optical spectroscopy is an excellent experimental
alternative to conventional spectroscopic methods that probe the low energy
electronic structure in strongly correlated electron systems \cite{Kabanov}.
In particular, it has been shown that carrier relaxation dynamics are very
sensitive to changes in the low energy density of states (e.g. associated
with the formation of a low energy gap or pseudogap) providing new insights
into the low energy electronic structure in these materials. In this report
we present the first studies of carrier relaxation dynamics in heavy fermion
(HF) systems by means of femtosecond time-resolved optical spectroscopy. Our
results show that the carrier relaxation dynamics, below the Kondo
temperature (T$_{K}$), are extremely sensitive to the low energy density of
states (DOS) near the Fermi level to which localized f-moments contribute.
Specifically, we have performed measurements of the photoinduced
reflectivity $\Delta $R/R dynamics as a function of temperature and
excitation intensity on the series of HF compounds YbXCu$_{4}$ (X=Ag, Cd,
In) \cite{Graf} in comparison to their non-magnetic counterparts LuXCu$_{4}$.

\section{Experimental}

The experiments were performed on freshly polished flux-grown single
crystals. We used a standard pump-probe set-up with a mode-locked
Ti:Sapphire laser (30 fs pulses at 800 nm and 80 MHz repetition rate) for
both pump and probe pulse trains. The photoinduced (PI) change in
reflectivity $\Delta $R/R was measured using photodiode and lock-in
detection. The pump laser fluence was kept below 10 $\mu $J/cm2 and the
pump/probe intensity ratio was $\sim $30. The steady-state heating effects
were accounted for as described in \cite{Mihailovic} yielding an uncertainty
in temperature of $\pm $3 K.

\section{Results and discussion}

In Fig. 1a we show the PI reflectivity traces as a function of temperature
on the HF system YbCdCu$_{4}$ (T$_{K}$ $\sim $100K, low temperature
Sommerfeld coefficient $\gamma $ $\sim $200 mJ/mol K$^{2}$). The relaxation
time is virtually temperature independent above $\ \sim $140 K and shows a
quasi-divergence (hundred-fold increase) as T = 0 K is approached.

Similar behavior (see Fig. 1b) is observed also in YbAgCu$_{4}$ (T$_{K}$ $%
\sim $100K, $\gamma $ $\sim $200 mJ/mol K%
%TCIMACRO{\UNICODE{0xb2}}%
%BeginExpansion
${{}^2}$%
%EndExpansion
) \cite{Graf}. Importantly, YbAgCu$_{4}$ is a much better metal than YbCdCu4
revealing that the observed dynamics are not merely due to changes in the
carrier concentration. Furthermore, the relaxation time of the non-HF
compound LuAgCu$_{4}$ behaves similar to regular metals \cite{Groenevald},
showing a gradual decrease upon cooling. This, coupled with the data on the
HF compounds in Fig. 1b, indicates that the upturn in the relaxation time at 
$\sim $T$_{K}$ and the subsequent divergence as T $\rightarrow $ 0 K derives
from the low energy electronic structure of these HF compounds.

\FRAME{fhFU}{6.8165in}{3.3667in}{0pt}{\Qcb{a) Normalized photoinduced
reflectivity in YbCdCu$_{4}$ at various temperatures above and below T$_{K}$ 
$\sim $100K. b) Temperature dependence of the relaxation time $\protect\tau %
_{R}$ for two different HF compounds, together with the power-law fits $%
\protect\tau _{R}=T^{-p}$, where p = 3 $\pm $ 0.5. Note that the dotted line
for LuAgCu4, which is not a HF compound, does not display this power law
behavior. c) Temperature dependence of the Sommerfeld constant $\protect%
\gamma $ for YbAgCu$_{4}$ as extracted from the relaxation time data (solid
squares) compared to the calculated $\protect\gamma $ for the case of a
single impurity model \protect\cite{Rajan} with the impurity spin 7/2.}}{}{%
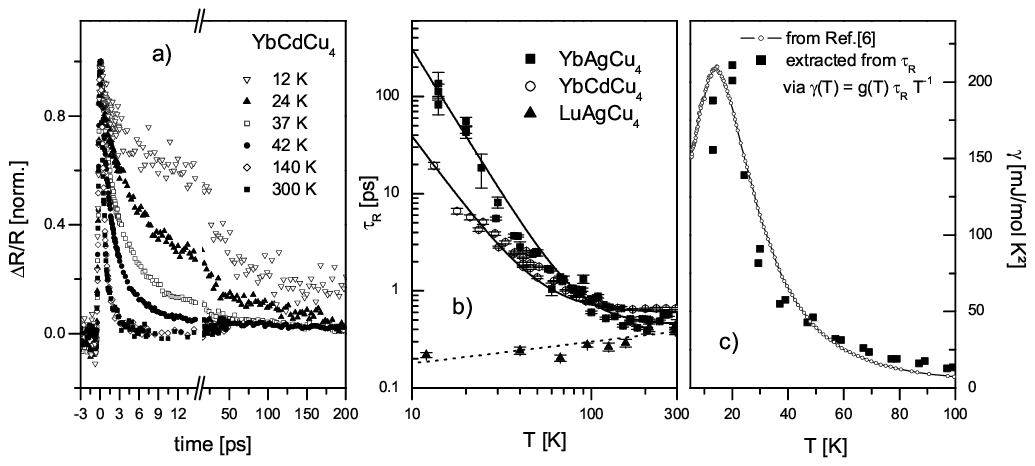}{\special{language "Scientific Word";type
"GRAPHIC";maintain-aspect-ratio TRUE;display "USEDEF";valid_file "F";width
6.8165in;height 3.3667in;depth 0pt;original-width 4.5264in;original-height
2.226in;cropleft "0";croptop "1";cropright "1";cropbottom "0";filename
'Figure.EPS';file-properties "XNPEU";}}

An important and unresolved issue in HF physics \cite{Degiorgi} is whether
the electronic ground state is a many-body Kondo state that forms only below
T$_{K}$, or is more simply described in terms of the thermal population of a
narrow gap system with the gap arising from hybridization between the
conduction electrons and localized \emph{f} electrons (note that the gap is
present at all temperatures above and below T$_{K}$ in this case). As we
have shown, the ultrafast dynamics are clearly sensitive to the low energy
electronic state and such experiments may provide insight into the true
nature of the HF ground state. However, a complete analysis (currently in
progress) requires many-body calculations beyond the scope of this paper. In
the following, we present a heuristic model, which provides some physical
insight into the low temperature electron dynamics in HF compounds.

The two-temperature model is commonly employed to interpret the ultrafast
dynamics in metals \cite{Groenevald} with the sequence of relaxation events
after photoexcitation as follows: the photoexcited carriers first rapidly
thermalize via electron-electron scattering ($\tau _{e-e}\approx $100 fs),
giving rise to a rapid change in reflectivity on a sub-picosecond timescale.
This is followed by energy transfer to the lattice with a characteristic
electron-phonon relaxation time $\tau _{e-ph}$ on the order of a picosecond 
\cite{Groenevald}. At low excitation fluences, as used here, the increase in
the electronic temperature after photoexcitation is small compared to the
lattice temperature, T. In this limit the T-dependence of $\tau _{e-ph}$ in
the two-temperature model is given by $\tau _{e-ph}=\gamma (T)T/g(T)$, where 
$g(T)$ is the electron-phonon coupling function \cite{Groenevald}. Since $%
g(T)$ is constant above the Debye temperature $\Theta _{D}$, and
proportional to T$^{4}$ at T\TEXTsymbol{<}\TEXTsymbol{<} $\Theta _{D}$, $%
\tau _{e-ph}$ is expected to decrease linearly upon cooling from room
temperature, showing a minimum with a subsequent increase going as T$^{-3}$
at low temperatures. However, this low-T upturn in $\tau _{R}$ is not
observed in LuAgCu$_{4}$ (Fig 1b) - as it was not observed in conventional
Fermi liquid metals like Au and Ag \cite{Groenevald}. The lack of the low-T
quasi-divergence in $\tau _{R}$ in conventional metals was shown to result
due to the fact that $\tau _{e-e}$ and $\tau _{e-ph}$ are comparable below $%
\approx $300-600 K \cite{Groenevald} in the experimental configuration of
low laser powers. A non-thermal electron model \cite{Groenevald} that takes
into account a ''finite'' $\tau _{e-e}$ can account for a) the increased
room temperature $\tau _{e-ph}$ and b) the low temperature plateau in the
observed temperature dependence of $\tau _{e-ph}$ in Au and Ag \cite
{Groenevald} as well as in LuAgCu$_{4}$.

However, in the HF compounds we have studied $\gamma $ is, below T$_{K}$,
two orders of magnitude larger than in conventional metals. This is expected
to give rise to a hundred-fold increase in $\tau _{e-ph}$ in these systems
and the two-temperature model (without considering the initial non-thermal
electron distribution) is expected to be valid. This would give rise to $%
\tau _{R}\propto $ $T^{-3}$ at low-T as experimentally observed (Fig. 1b).
Furthermore, since $g(T)$ depends only on $\Theta _{D}$ and a prefactor that
can be estimated from the high-T value of $\tau _{R}$ \cite{Groenevald} (at
300K the value of $\tau _{R}$ is close to the estimated $\tau _{e-ph}$), one
can extract the Sommerfeld constant via $\gamma (T)=g(T)\tau _{R}/T$. We
find a remarkable agreement between the low temperature $\gamma (T)$
extracted in this way from $\tau _{R}(T)$ and the theoretical calculation 
\cite{Rajan} for the case of single impurity model with impurity spin 7/2 -
Fig 1c, giving further support to this analysis.

\section{Conclusions}

In conclusion, we have utilized ultrafast optical techniques to study the
dynamics of quasiparticles in HF compounds YbXCu$_{4}$. We have observed a
power-law divergence in the relaxation time at low temperatures. The
dramatic hundred-fold increase in the relaxation time at low temperatures in
YbXCu$_{4}$ (and the lack of this quasi-divergence in the non-HF LuXCu$_{4}$
analogs) clearly demonstrates that our measurements probe the heavy fermion
ground state.

\bigskip \bigskip 

* phone: 1-505-665-8839, fax 1-505-665-7652, e-mail: jdemsar@lanl.gov


\begin{thebibliography}{9}
\bibitem{Kabanov}  V.V.Kabanov et al., Phys.Rev.B 59, (1999) 1497, J.Demsar,
D.Mihailovic, K.Biljakovic, Phys.Rev.Lett. 83, 800 (1999).

\bibitem{Graf}  T.Graf et al., Phys. Rev. B 51 (1995) 15053; J. L. Sarrao et
al., Phys. Rev. B 59 (1999) 6855.

\bibitem{Mihailovic}  D. Mihailovic and J. Demsar in Spectrosopy of
Superconducting Materials, ed. Eric Falques, ACS Symposium Series 730; The
American Chemical Society: Washington, D.C., 1999.

\bibitem{Groenevald}  R. H. M. Groeneveld, R. Sprik, A. Lagendijk, Phys.
Rev. B 51 (1995) 11433, and the references therein.

\bibitem{Degiorgi}  L. Degiorgi, Rev. Mod. Phys. 71, 687 (1999) and the
references therein.

\bibitem{Rajan}  V.T.Rajan, Phys.Rev.Lett.51 (1983) 308.
\end{thebibliography}
\end{document}